\documentstyle[aps,prl,multicol,epsfig,amsbsy,amssymb,amstex]{revtex}

\def\bbbc{{\mathchoice
{\setbox0=\hbox{$\displaystyle\rm C$}\hbox{\hbox
to0pt{\kern0.4\wd0\vrule height0.9\ht0\hss}\box0}}
{\setbox0=\hbox{$\textstyle\rm C$}\hbox{\hbox
to0pt{\kern0.4\wd0\vrule height0.9\ht0\hss}\box0}}
{\setbox0=\hbox{$\scriptstyle\rm C$}\hbox{\hbox
to0pt{\kern0.4\wd0\vrule height0.9\ht0\hss}\box0}}
{\setbox0=\hbox{$\scriptscriptstyle\rm C$}\hbox{\hbox
to0pt{\kern0.4\wd0\vrule height0.9\ht0\hss}\box0}}}}

\newcommand{\beq}{\begin{eqnarray}}
\newcommand{\eeq}{\end{eqnarray}}

\begin{document}
\title{Impurity states and interlayer tunneling in  high temperature
superconductors}
\author{I. Martin$^1$, A.V. Balatsky$^1$, and J. Zaanen$^2$}
\address{$^1$ Theoretical Division, Los Alamos
National Laboratory, Los Alamos, NM 87545\\
$^2$ Leiden Institute of Physics, Leiden University, 2333CA
Leiden, Netherlands}

\date{Printed \today }

\maketitle

\begin{abstract}
We argue that the Scanning Tunneling Microscope  (STM) images of
resonant states generated by doping Zn or Ni impurities into Cu-O
planes of BSCCO are the result of  quantum interference of the
impurity signal coming from several distinct paths. The impurity
image seen on the surface is greatly affected by interlayer
tunneling matrix elements. We find that the optimal tunneling path
between the STM tip and the metal (Cu, Zn, or Ni) $d_{x^2 - y^2}$
orbitals in the Cu-O plane involves intermediate excited states.
This tunneling path leads to the four-fold nonlocal filter of the
impurity state in Cu-O plane that explains the experimental
impurity spectra.  Applications of the tunneling filter to the
Cu vacancy defects and ``direct'' tunneling into Cu-O planes are also
discussed.

\end{abstract}
\pacs{Pacs Numbers: XXXXXXXXX}

\vspace*{-0.4cm}
\begin{multicols}{2}

\columnseprule 0pt

\narrowtext
\vspace*{-0.5cm}

Recently J.C. Davis and collaborators applied the STM technique to
image single Zn and Ni impurities in optimally doped
BSCCO\cite{Pan1,Pan2}.  These experiments proved that one can image
single impurity states in an unconventional superconductor and
demonstrated the highly anisotropic structure of these states.

Although it appears that on a gross scale these findings can be
understood in terms of a conventional $d$-wave superconductor
perturbed by potential scattering, upon closer inspection problems of
principle seem to arise.  The impurity states observed by STM are
characterized by two main features: (1) energy and width of the
impurity-induced resonance in the density of states (DOS), and (2)
the spatial structure of the resonance.   While the DOS seems to be
satisfactorily described by a single-site impurity model
\cite{balatsky}, the real space distribution of intensity cannot be
fit by this model.  The main problem with the Zn impurity image seen
in the STM experiments is that the intensity of the signal on the
impurity site is very bright, which is at odds with the unitary
scattering off Zn.  We remind that Zn$^{2+}$ has a closed $d$ shell
and hence, exactly on the impurity site the scattering potential is
very strong.  Unitary scattering is equivalent to the hard wall
condition for the conduction states and therefore {\em no or very
little} intensity of electron states is expected on the Zn site. A
similar problem arises also with explaining the Ni-induced resonance.

Here we demonstrate that these problems find a natural resolution
in terms of the specific way in which the local density of states
of the cuprate planes is probed in the STM experiments. We argue
that the quantum-mechanical nature of the tunneling from the STM
tip into the Cu-O layer that hosts impurity requires tunneling
through the uppermost insulating Bi-O layer which effectively
filters the signal.  Surprisingly, similar filtering should also
take place even in the case of ``direct'' tunneling into Cu-O
plane. Such non-local tunneling has profound consequences for the
real space image of the impurity state seen by STM.

\begin{figure}[htbp]
\begin{center}
\includegraphics[width = 3.0 in, height=2 in]{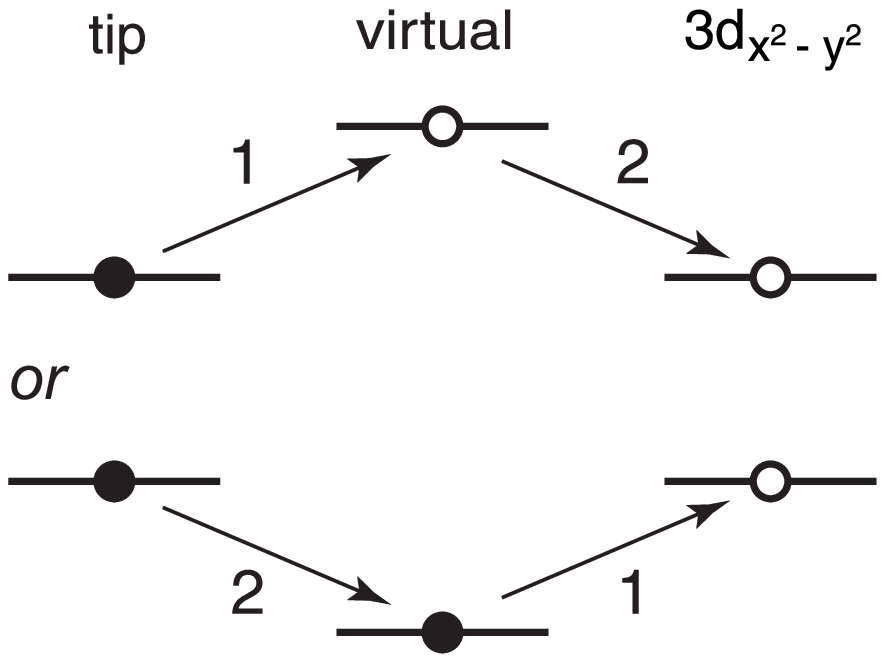}\\
(a)\\
\vspace{0.3cm}
\includegraphics[width = 3.0 in, height=1.8 in]{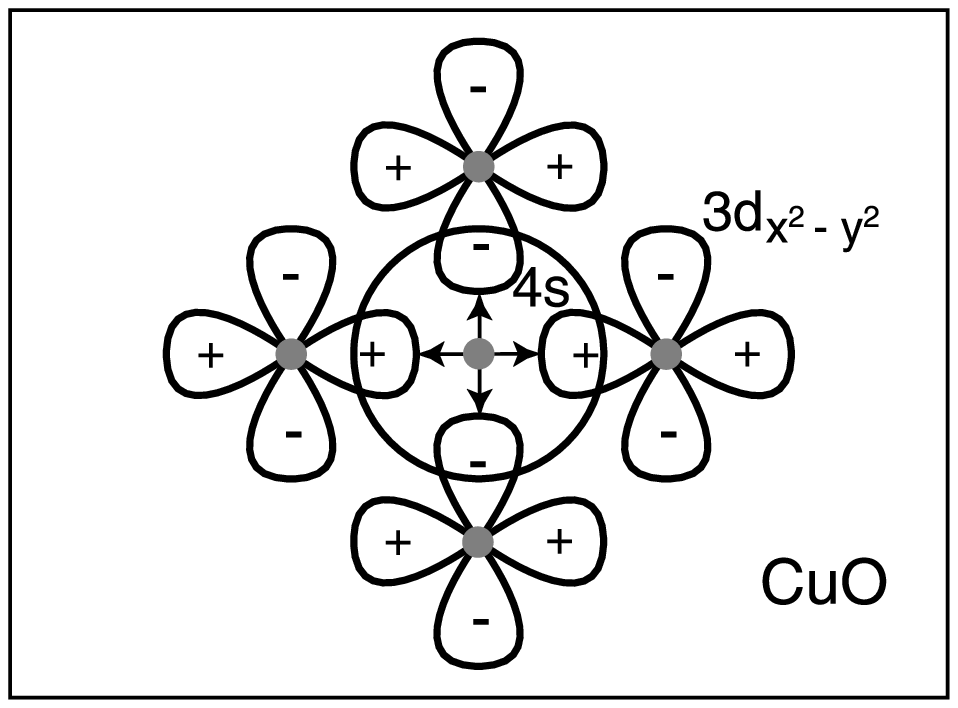}\\
(b)\\
\vspace{0.3cm}

\caption{(a) Tunneling from STM tip into $3 d_{x^2-y^2}$
orbitals in the Cu-O layer occurs through the virtual energy
states of Bi and Cu. The tunneling processes involve both empty
(top) and occupied (bottom) intermediate states.  All such
processes contribute coherently to the tunneling amplitude from
the tip to the $3 d_{x^2-y^2}$ orbitals. (b) Tunneling through a
particular Bi atom {\em does not} probe the $3 d_{x^2-y^2}$
orbital of the metal atom (Cu, Zn or Ni) right underneath it due
to the vanishing overlap. Instead, the tunneling involves
orbitals that extend out of the planes, such as $4s$ (shown) or
$3d_{3z^2 - r^2}$ of Cu and $6p_z$ of Bi.   These orbitals are
symmetric in the Cu-O plane and hence couple to the {\em
neighboring} metal $3 d_{x^2-y^2}$ orbitals through the
$d$-wave-like $fork$, Eq.~(\protect\ref{eq:filter}).} 
\label{fig:fork} 
\end{center}
\end{figure}

There are two major types of the tunneling routes between the
STM tip and the conducting orbitals in the Cu-O plane: (a) {\em
Direct} tunneling due to the overlap between the tip and the
planar $3d_{x^2-y^2}$ wavefunctions, and (b) {\em indirect}
tunneling through intermediate excited (occupied or empty)
states, Fig.~\ref{fig:fork}a.  The direct tunneling probability
over the experimentally relevant distances of about 10\AA
\cite{Pan1,Pan2} is however exponentially small\cite{tersoff}. 
On the other hand, the importance of the excited states for
mediating STM tunneling follows directly from the Bi-O
topographs that clearly show the positions of the nominally
insulating Bi atoms on the surface of BSCCO. The analysis of the
topographs implies that the excited Bi orbitals {\em focus} the
flow of the tunneling electrons.  Therefore, we argue that the
indirect tunneling via overlapping intermediate orbitals is the
dominant tunneling mechanism.

\begin{figure}[htbp]
\begin{center}
\includegraphics[width = 3.4 in]{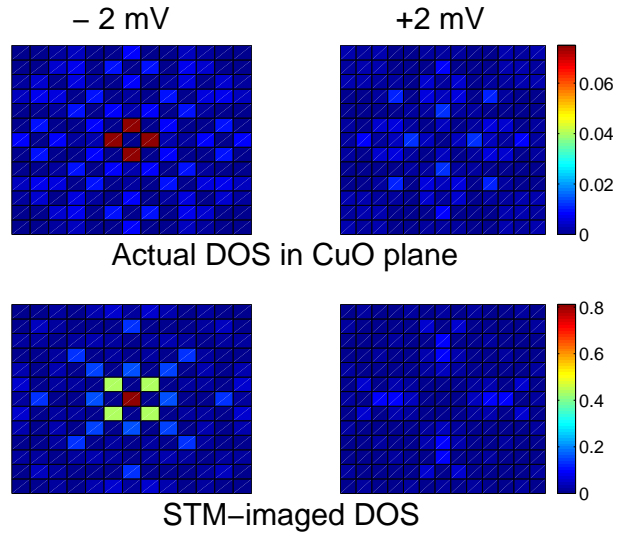}
\vspace{0.5cm}

\caption{Top two plots show the real-space spectral intensity of
the calculated Zn impurity  state in the Cu-O plane.  While most
of the intensity is concentrated at the negative bias, -2 meV, as
it is seen in the experiments \protect\cite{Pan1}, the  spatial
shape of the state does not agree with the experimental results.
The bottom two plots show the same impurity state, but as seen
through the Bi-O layer.  The  figure is obtained by applying the
filter function of Eq.~(\protect\ref{eq:filter}).  Both the energy
of the state and the spatial intensity distribution agree with
experiment \protect\cite{Pan1}.} \label{fig:RI}
\end{center}
\end{figure}

\begin{figure}[htbp]
\begin{center}
\includegraphics[width = 3.4 in]{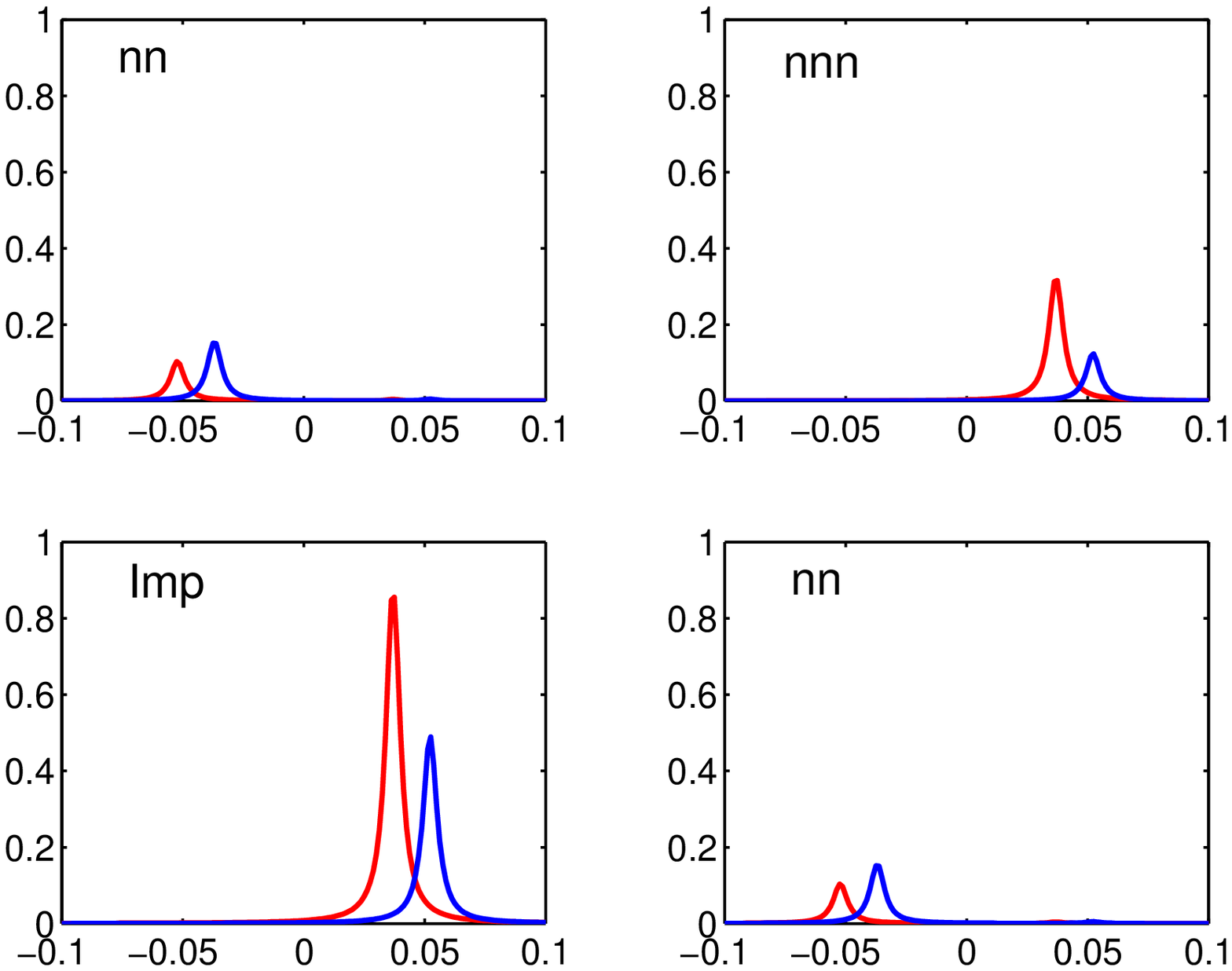}\\
{\large (a)}\\
\includegraphics[width = 3.4 in]{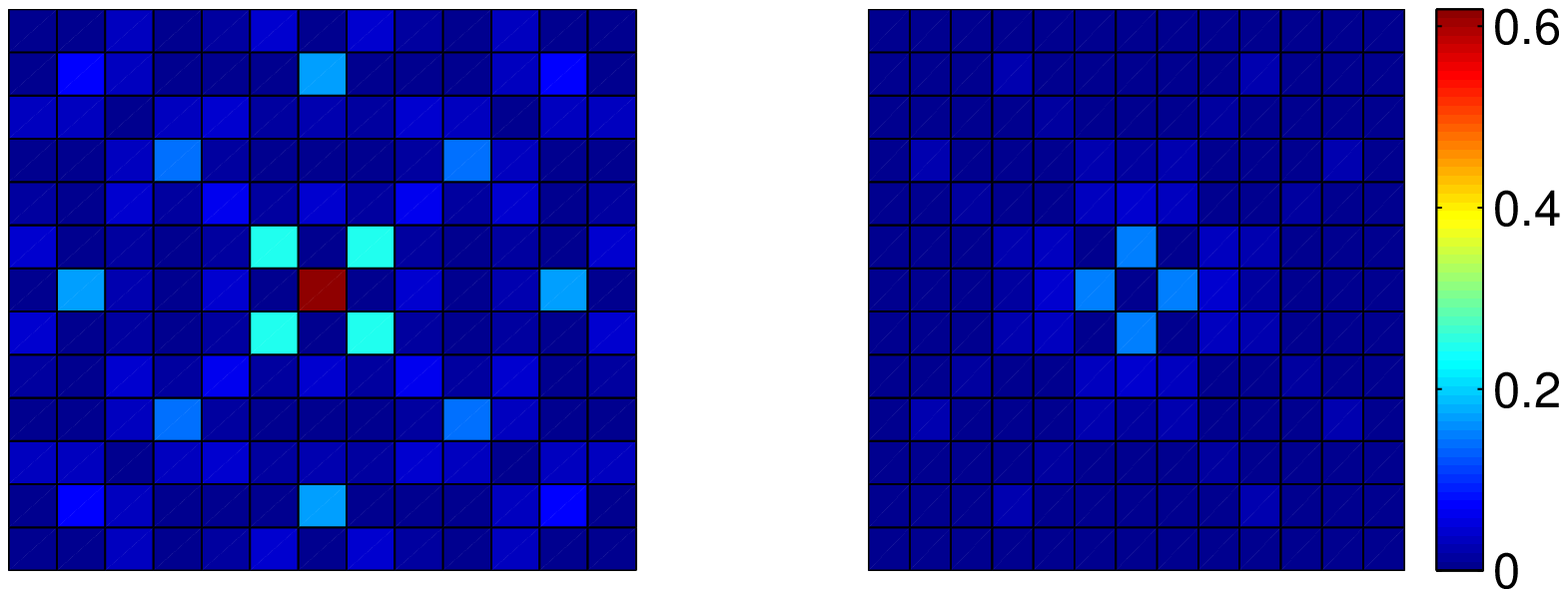}\\
{\large (b)}

\vspace{0.5cm}

\caption{Simulation of Ni as an impurity with weak mixed
potential, $V_{\rm Ni} = t$ and $S_{\rm Ni} = 0.4 t$. (a) Spectral
intensity on the impurity site, its nearest neighbors (nn), and
next nearest neighbor (nnn).  Red line is spin up, and blue is
spin down. (b) Intensity map for the combined spin up and spin
down states.} \label{fig:Ni}
\end{center}
\end{figure}

The strongest indirect tunneling channel involves intermediate
states that have the largest overlaps. In the case of BSCCO, these
are the orbitals that extend out of the planes, such as $4s$ or
$3d_{3z^2-r^2}$ of Cu and 6$p_z$ of Bi.  Furthermore, only the
states with zero in-plane orbital momentum have non-zero overlap
with the apical oxygen 2$p_z$ and 3$s$ orbitals that play
important role in the interlayer communication.  Being radially
symmetric in the Cu-O plane, such states {\em cannot} couple to
the relevant $3 d_{x^2-y^2}$ states {\em on-site}.  They do couple
however to the {\em neighboring} $3d_{x^2-y^2}$ orbitals through
the $d$-wave-like $fork$ (Fig.~\ref{fig:fork}b).  The resulting
tunneling amplitude is
\begin{equation}
\label{eq:filter} M_{i,j}\sim \Psi_{i+1,j} + \Psi_{i-1,j} -
\Psi_{i,j+1} - \Psi_{i,j-1},
\end{equation}
where $\Psi_{i,j}$ is the impurity state wavefunction on site
$(i,j)$.  Hence, the symmetry analysis leads to the surprising
conclusion: Tunneling on top of a particular Cu (or Zn or Ni) atom
in the Cu-O plane {\em does not} probe its $3d_{x^2-y^2}$
orbitals, but rather measures a linear combination of its
neighbors.  The tunneling matrix element determines the intensity
of the impurity signal $F[{\Psi_{i,j}}] = |M_{i,j}|^2$.
Calculating the spectral intensities in the vicinity of the
impurity using this filtering function reproduces the features
seen in the experiments in great detail (Figs.~\ref{fig:RI} and
\ref{fig:Ni}). The aforementioned anomalies find an explanation in
terms of interferences associated with the non-local way in which
the electronic states are probed. The described filtering
mechanism is related to the one responsible for the interlayer
tunneling in the bulk cuprates \cite{chak,oka}.  An alternative
non-local filter based on the incoherent {\em direct} tunneling
from the tip  into Cu-O plane was recently studied by Zhu {\em et
al.}\cite{zhu}.  The angular effect of the filter on the far
asymptotic of the impurity states was discussed earlier
\cite{Bal1}.

Similar to the Zn case, the Ni impurity tunneling intensity is
maximal at the Bi position immediately above the impurity site.
However, there are several important differences between the
observed  Zn- and Ni-induced states: (1) The energy of the Zn
state is close to the chemical potential,  $\varepsilon_{\rm
Zn}= -2$ mV, while the Ni state energy is larger and  is split,
$\varepsilon_{\rm Ni}= 9\ {\rm and}\ 18$ mV, (2)  the Zn state
appears  only on the negative bias, while the Ni state shows up
both on positive bias and  the symmetric negative bias.  In this
letter we demonstrate that these experimental features can also
be reproduced within the standard theory of the impurity
states\cite{balatsky}, with the spatial structure of the states
being reproduced by properly taking into account the {\em fork}
effect.

The starting point of our model is a two-dimensional mean-field
(MF) Hamiltonian with the nearest neighbor attraction, $V$, which
yields $d$-wave superconductivity in the range of dopings close to
half-filling, \beq\label{eq:H_MF} H_0 = -
\sum_{i,j,\sigma}{t_{ij}c^\dagger_{i\sigma} c^{\;}_{j\sigma}} +
\sum_{\langle ij \rangle}{c^{\;}_{i\downarrow} c^{\;}_{j\uparrow}
\Delta_{ij}^* + {\rm h.c.}} \ . \eeq Here, $\Delta_{ij} = V
\langle c_{i\downarrow} c_{j\uparrow}\rangle$ is  the
self-consistent MF superconducting order parameter.  The hopping
$t_{ij}$ equals $t$ for nearest neighbors and $t^\prime$ for for
the  second-nearest neighbor sites $i$ and $j$.  The parameters
that are relevant for BSCCO  are $t = 400$ meV, $t^{\prime} = -
0.3 t$.  To match the amplitude of the superconducting gap in
optimally-doped BSCCO, which is about 40 meV, we choose the
attraction $V = - 0.525 t$.  The chemical potential is chosen to
yield 16\% doping ($\mu = -t$).

The local impurity is introduced into Hamiltonian Eq.~(\ref{eq:H_MF}) by
modifying the electron energy on a particular
site\cite{YuLu,Yazdani,byers,balatsky}.  The corresponding addition to the
Hamiltonian is
\beq\label{eq:Himp}
H_{\rm imp} = V_{\rm imp}(n_{0\uparrow} +
n_{0\downarrow}) + S_{\rm imp}(n_{0\uparrow} -
n_{0\downarrow}).
\eeq
The first term is the potential part of the impurity energy that  couples to
the total electronic density on site 0, and the second term describes the
magnetic interaction of the impurity spin and the electronic spin density on
the same site.  We assume that the impurity spin is large and can be  treated
classically, as if it were a local magnetic field. The goal is to
determine $V_{\rm imp}$ and $S_{\rm imp}$ so as  to match both the location of
the impurity states within the gap and the spatial  distribution of their
intensity.

\begin{figure}[htbp]
\begin{center}
\includegraphics[width = 3.0 in]{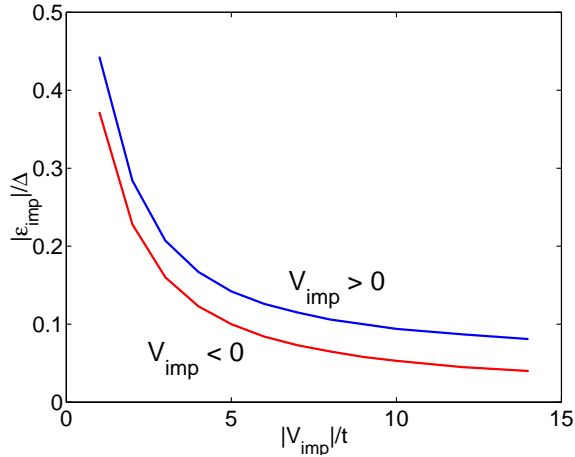}
\vspace{0.5cm}

\caption{Impurity energy level position as a function of the potential
impurity strength, $V_{\rm imp}$.  The blue (top) line corresponds to
repulsive impurities, which produce positive energy levels
($\varepsilon_{\rm  imp} > 0$); the red line is for attractive impurities,
which generate negative  energy states ($\varepsilon_{\rm imp} < 0$).  For
strong impurity potential  both energies converge to the same small value
determined by the amount of particle-hole symmetry breaking.}
\label{fig:Elev}
\end{center}
\end{figure}

First let us analyze the position of the impurity energy level as
a  function of the impurity potential.  We solve the MF equations
self-consistently on a square lattice with periodic boundary
conditions.  The results are presented in Figure \ref{fig:Elev}.
All impurities placed in a superconductor generate quasiparticle
weight both on positive and symmetric negative biases. The sign of
the  energy level is defined  based on where the majority of the
quasiparticle  weight resides. Attractive impurities produce
energy levels lying below the Fermi surface, while repulsive ones
generate states with positive energy.   In the limit of a very
strong impurity potential both attractive and repulsive impurities
generate identical states with a small residual energy  related to
the amount of the particle-hole symmetry breaking, caused by the
specifics  of the band structure and doping \cite{kruis}.

The analysis of the energy level positions implies that the Zn
impurity can be associated with a strong attractive potential,
$V_{\rm Zn} = - 11 t = -4.4$ eV and $\varepsilon_{\rm Zn} = -
0.005 t = - 2$ meV.  Ignoring for now the level splitting, the Ni
case can be associated with a relatively weak  repulsive impurity,
$V_{\rm Ni} = t$ and $\varepsilon_{\rm Ni} = 0.0443 t = 18$  meV.
These impurity strengths are in agreement with the general band
structure arguments.  Zn$^{2+}$ ion has 10 electrons that
completely fill $d$ orbitals. Hence, the $d_{x^2-y^2}$ orbital of
Zn, relevant for interaction with Cu-O plane orbitals,  is deep
below the Fermi surface.  On the other hand, Ni$^{2+}$ has 8
electrons in the $d$ shell, with the  $d_{x^2-y^2}$ being
unoccupied, but with a small energy, given by the level splitting
within the $d$ shell.

The spatial distribution of the spectral intensity corresponding
to the Zn impurity is shown in Figure~\ref{fig:RI}.  The top two
plots show the intensity $A_{i,j}^{\rm CuO}$ as it would be seen
if STM tip were directly imaging DOS in the Cu-O layer.   The
intensity is related to impurity state wavefunction $\Psi$  on
site $(i,j)$,
\beq A_{i,j}^{\rm CuO} =|\Psi_{i,j}|^2. \eeq
The intensity on the impurity site is suppressed due to the  strong impurity
potential. The bottom two plots correspond to imaging through the top Bi-O
layer.  They are  obtained by applying a filtering function $F[\Psi] =
|M_{i,j}|^2$ to the impurity state wavefunction $\Psi$. The effect of the
filtering function is to produce the intensity
\beq A_{i,j}^{\rm BiO} \propto|\Psi_{i+1,j}+\Psi_{i-1,j}-\Psi_{i,j+1}-
\Psi_{i,j-1}|^2.
\eeq
Indeed, the intensity on the Bi-O layer is maximized on the impurity site due
to the interference of the contributions from the impurity's nearest neighbor
sites.

The structure of the Ni-induced  state is more complicated than
the Zn case.  It appears on both positive and negative biases. In
addition, there is a peak splitting on  each bias.  Unlike
Zn$^{2+}$, Ni$^{2+}$ impurity is magnetic, with spin 1. To
simulate the effect of spin we include  a non-zero magnetic part
of the impurity potential, $S_{\rm imp} \ne 0$, in the Hamiltonian
Eq.~(\ref{eq:Himp}).   The spin  component introduces level
splitting between ``up'' and ``down'' spin states.
Figure~\ref{fig:Ni}a shows the amplitudes of the spin-split states
on the impurity sites and its neighbors for $V_{\rm Ni} = t$ and
$S_{\rm Ni} = 0.4 t$.  The spin-split energy levels are
$\varepsilon_{\downarrow} =  0.052 t$ and $\varepsilon_{\uparrow}
= 0.037 t$. The total spin-up and spin-down intensities for each
bias are shown in Figure~\ref{fig:Ni}b.  Both  figures (a) and (b)
correspond to the image affected by the fork filter.  The general
shape of the states agrees well with the experimental
data\cite{Pan2}. The total weight on the  negative bias is about
factor of 3 smaller than the weight on the positive bias.

Some aspects of the Ni-impurity state cannot be addressed within
the  simple framework of our theory.  We assumed that the spin
of the impurity is static and plays the role of a local magnetic
field.   Bulk measurements  indicate that Ni dopants do not
introduce additional magnetic moments.  This may be either due
to the partial Kondo screening of $S=1$ Ni spin or due to the
local attraction of additional hole (``self-doping'') by
Ni$^{2+}$ to form an effective Ni$^{3+}$ complex with spin 1/2. 
No magnetic signature, combined with the weak potential
character of Ni may explain the relatively weak pair-braking
effect of Ni doping observed experimentally.

In the case of Zn, we focused on the large {\em local} potential
scattering off the impurity as a main effect to account for the
STM intensity distribution. It is important to mention that Zn
substitution also induces local magnetic moment, which is observed
in magnetic susceptibility and NMR measurements \cite{YBCO}.  The
quenching of magnetic moment around Zn can lead to the Kondo
resonance near the Fermi surface, as has been proposed by
Polkovnikov {\em et al.} \cite{sachdev}. Below the Kondo temperature,
Zn should behave as a unitary scatterer, virtually
indistinguishable from either local or extended\cite{sachdev,zhu2}
potential scatterer.  We argue however, that the tunneling fork
filter should apply regardless of the details of the scattering
mechanism.

The tunneling fork mechanism has consequences beyond the
settings of the original experiments \cite{Pan1,Pan2}.  Here we
point out two of them: (1) the direct tunneling into Cu-O plane,
and (2) the structure of the Cu vacancy states.  One could
assume that the tunneling into the exposed top Cu-O plane should
be free of the tunneling fork. However, upon closer inspection
it is clear that even in this case, the direct tunneling from
tip into the planar 3$d_{x^2-y^2}$ orbitals is exponentially
weak compared to the indirect tunneling through the s-wave like
orbitals of Cu, Zn, or Ni extending out of the plane.  Hence,
the tunneling fork should apply also in this case, resulting in
the same spatial form of the filtered impurity states as
obtained above.   In the case of Cu vacancy in the Cu-O plane,
under assumption that there are no $s$-wave-like orbitals
centered at the vacancy site, the tunneling fork mechanism
implies that resonance state observed by STM should be the same
as for Zn, but with no spectral intensity in the center of the
pattern.

In conclusion, we have demonstrated that recently observed Zn
and Ni impurity states in BSCCO\cite{Pan1,Pan2} can be explained
by simple model of strong potential  impurity in the case of Zn
and mixed (potential + spin) impurity in the case of Ni,
interacting with a $d$-wave superconducting condensate.  The
crucial  aspect that we have included in the present treatment
is the effect of the quantum-mechanical tunneling between the
STM tip and the Cu-O planes.  The same non-local filter effect
should be operable in the case of the ``direct'' tunneling into
Cu-O planes and should have consequences for other types of
defects, e.g. Cu vacancies.

We are grateful to J.C. Davis, S.H. Pan, S. Sachdev, M. Vojta,
and J.X. Zhu for useful discussions. This work has been
supported by the U.S. DoE.


\end{multicols}
\end{document}